\documentclass[10pt,journal,twocolumn,twoside]{IEEEtran} 
\usepackage{xpatch}
\usepackage{graphicx}
\usepackage{epstopdf}
\usepackage{float}
\usepackage{algorithmic}
\usepackage{array}
\usepackage{amsmath}
\usepackage{amssymb}
\usepackage{mdwmath}
\usepackage{mdwtab}
\usepackage{eqparbox}
\usepackage{stfloats}
\usepackage{fixltx2e}
\usepackage{tabularx}
\usepackage{cases} 

\usepackage{flushend}

\usepackage{xcolor}
\usepackage{makecell}

\usepackage[boxed,ruled,commentsnumbered]{algorithm2e}
\usepackage{url}
\usepackage{cite}
\ifCLASSOPTIONcompsoc
\usepackage[caption=false,font=normalsize,labelfont=sf,textfont=sf]{subfig}
\else
\usepackage[caption=false,font=footnotesize]{subfig}
\fi
\allowdisplaybreaks[4]

\makeatletter

\newcommand{\Rmnum}[1]{\expandafter\@slowromancap\romannumeral #1@}
\makeatother

\makeatletter
\renewcommand*{\@opargbegintheorem}[3]{\trivlist
      \item[\hskip \labelsep{\bfseries #1\ #2}] \textbf{(#3):}\ }
\makeatother

\begin{document}

\makeatletter
\def\changeBibColor#1{%
  \in@{#1}{}
  \ifin@\color{red}\else\normalcolor\fi
}
 
\xpatchcmd\@bibitem
  {\item}
  {\changeBibColor{#1}\item}
  {}{\fail}
 
\xpatchcmd\@lbibitem
  {\item}
  {\changeBibColor{#2}\item}
  {}{\fail}
\makeatother

\title
{Look-Ahead Task Offloading for Multi-User Mobile Augmented Reality in Edge-Cloud Computing}
\author{Ruxiao Chen, and Shuaishuai Guo,~\IEEEmembership{Senior Member, IEEE} 
\thanks{The work is supported in part by the National Natural Science Foundation of China under Grant 62171262; in part by Shandong Provincial Natural Science Foundation under Grant ZR2021YQ47; in part by the Taishan Young Scholar under Grant tsqn201909043; in part by Major Scientific and Technological Innovation Project of Shandong Province under Grant 2020CXGC010109. (\emph{*Corresponding author: Shuaishuai Guo}).}
\thanks{R. Chen and S. Guo are all  with the School of Control Science and Engineering, Shandong University, China. S. Guo is also with Shandong Key Laboratory of Wireless Communication Technologies, Shandong University, China (e-mail: ruxiaochen333@gmail.com; shuaishuai$\_$guo@sdu.edu.cn). }
   }
\maketitle


\begin{abstract} 
Mobile augmented reality (MAR) blends a real scenario with overlaid virtual content, which has been envisioned as one of the ubiquitous interfaces to the Metaverse. Due to the limited computing power and battery life of MAR devices, it is common to offload the computation tasks to edge or cloud servers in close proximity. However, existing offloading solutions developed for MAR tasks suffer from high migration overhead, poor scalability, and short-sightedness when applied in provisioning multi-user MAR services. 
To address these issues, a MAR service-oriented task offloading scheme is designed and evaluated in edge-cloud computing networks. Specifically, the task interdependency of MAR applications is firstly analyzed and modeled  by using directed acyclic graphs. 
Then, we propose a look-ahead  offloading scheme based on a modified Monte Carlo tree (MMCT) search, which can run several multi-step executions in advance to get an estimate of the long-term effect of immediate action.
Experiment results show that the proposed offloading scheme can effectively improve the quality of service (QoS) in provisioning multi-user MAR services, compared to  four benchmark schemes. Furthermore, it is also shown that the proposed solution is stable and suitable for applications in a highly volatile environment. 

\end{abstract}


\section{Introduction} 
\IEEEPARstart{M}{obile} augmented reality (MAR) has been widely applied in various fields such as gaming and education, as it can combine digital content with the physical world in real-time.  With the development of the fifth generation (5G) communication networks and edge-cloud computing, MAR is gaining increasing attention from both industry and academia. It has been envisioned as one of the ubiquitous interfaces to the Metaverse.
In pursuit of a high-quality user experience, MAR generally runs on portable devices such as mobile phones, AR headsets, and AR glasses. However, MAR tasks’ requirements on latency, mobility, and endurance are stringent and the limited computation capability and battery life of mobile devices greatly hinder its wide applications. To solve this problem, an edge-cloud computing paradigm has been applied in this field. It is acknowledged that MAR can be divided into five interdependent components: a video capturer, a feature extractor, a mapper, a tracker, an object recognizer, and a render. Among these components, the video capturer and render parts directly interact with users, and thus can only be run locally, while the other four components can be offloaded.
The basic idea of this paradigm is to selectively offload the computation tasks toward nearby edge or cloud servers with considerable computation power, thus is suitable for MAR devices to overcome its low battery capacity and computation ability. The implementation of this paradigm in MAR scenario has gained the interest of many researchers, but also raises many questions.

Several artificial intelligence-based algorithms like deep reinforcement learning (DRL) and heuristic algorithms like genetic algorithms (GA) and particle swarm optimization (PSO) were proposed for offloading MAR tasks in edge-cloud computing networks. For example,  Chen \emph{et al.} modeled offloading decision-making as a joint optimization problem and proposed a DRL-based algorithm using a multiagent deep deterministic policy gradient (MADDPG) framework, which minimized the energy consumption under the constraints of latency requirements \cite{Chen2021}. 
In \cite{Al-Shuwaili2017}, Shuwaili \emph{et al.} leveraged the inherent collaborative nature of MAR that mobile devices connected to the same base station have partly shared inputs and outputs to avoid extra computing, then using a successive convex approximation method (SCA) to solve the non-convex optimization problem.  Later, Brand \emph{et al.} extended the traditional single-path edge offloading model to a multi-path model with extra choices including device-to-device, edge, and cloud offloading in \cite{BRAUD2020}, which decreases the instability of mobile devices compared to single-path offloading.  In \cite{Wang2020}, Wang \emph{et al.} proposed a scheme named Closure, which has been shown to efficiently manage heterogeneous devices by calculating the Nash equilibrium of the attack-defense game model.

However, the aforementioned works schedule a single task at a time and only take into account the immediate effect of a decision without considering its long-term effect and the interdependency between different MAR components. Consequently, a large number of non-local jumps in the search space will be generated, leading to additional migration overheads. When an unfinished task is transferred from one host to another, it will cost extra transmission overhead and disrupt the original computing process of that host. Besides, it is noteworthy that the DRL-based and heuristic methods for offloading decision-making are poorly scalable. As the number of users grows, the complexity increases dramatically, requiring considerable iterations to converge to a new effective model or a relatively optimal solution, which consumes a large amount of time and energy. The open research direction of this field is to find a task offloading method that can reduce the migration overhead and overcome short-sightedness. The corresponding challenging issues are how to develop an objective function that efficiently models the migration overheads, what features should be used for offloading possess, and how to simulate the migration overheads.

Most recently, a workflow scheduling scheme named Monte Carlo tree search with a deep surrogate model (MCDS) algorithm  was proposed by Shreshth Tuli \emph{et.al} in\cite{Tuli2022a}. It is a look-ahead scheme that runs several multi-step executions in advance, which seems to be a solution to the existing problems in MAR offloading. However, its training process is not designed specifically for AR applications, and it uses the traditional Monte Carlo tree search to simulate the overhead of the process with no mechanism to narrow the search scope and to terminate in advance, resulting in a large amount of unnecessary computing resource and time loss.
\begin{figure*}
  \centering
  \includegraphics[width=16.5cm]{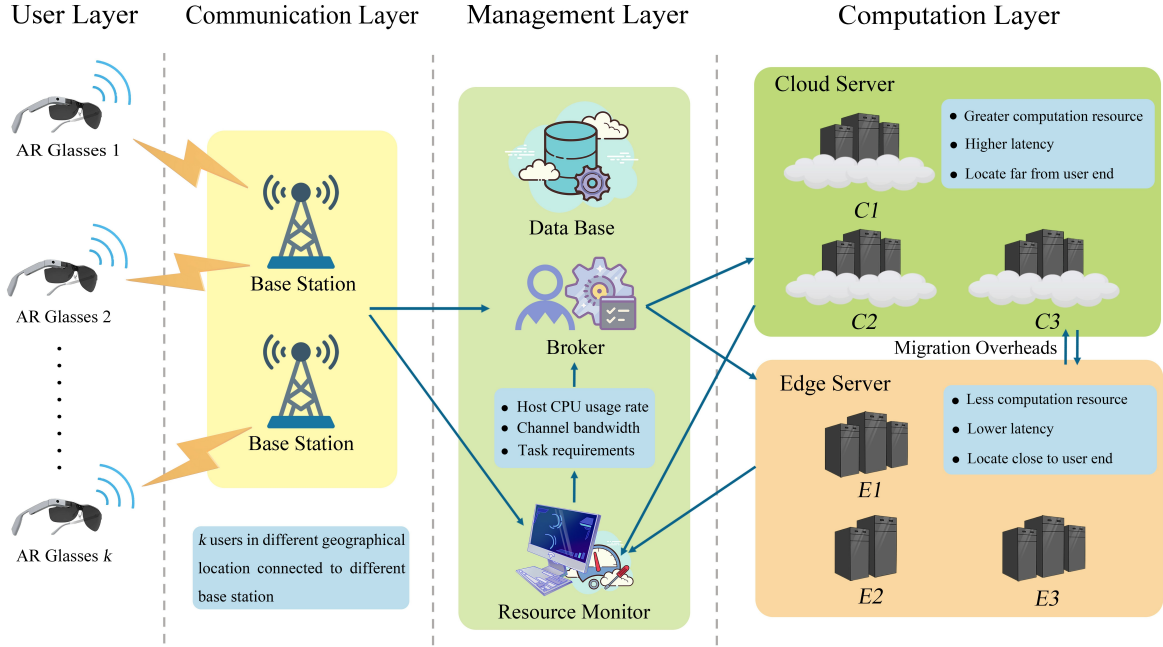}
  \caption{System architecture with four layers: user layer, communication layer, management layer, and computation layer.} 
  \label{fig1}
\end{figure*}
Drawing all the insights above, this article proposed a look-ahead offloading method using a modified Monte Carlo tree (MMCT) search in edge-cloud computing systems. The main contributions of this article can be summarized as follow:
\begin{itemize}
	\item A more specific and refined MAR application model is established using three types of directed acyclic graphs (DAG)  according to the MAR task interdependency.
	\item A long-sighted scheduling scheme based on MMCT is formulated, which is able to make offloading decisions with consideration of migration overhead and long-term QoS in real-time.
	\item The offloading method is applied in an edge-cloud collaborative computing system with multiple offloading paths. Experimental results show that it significantly reduces the influence of changing latency of geographically distributed mobile devices. 
\end{itemize}


\section{System Architecture}

Considering the computational heterogeneity and mobility of MAR tasks, we study an edge-cloud collaborative system with multiple edge and cloud servers. The system provides multiple offloading paths, which makes task upload and execution latency more stable\cite{BRAUD2020}. The edge-cloud system can be divided into four layers: a user layer, a communication layer, a management layer, and a computation layer. The overall structure is shown in Fig. \ref{fig1}. 

As shown in Fig. 1, we consider there are $k$ users distributed in different geographical locations in the user layer, each of which is connected to a base station in the communication layer according to its location. The management layer is an abstract layer that exists in the form of software, consisting of resource monitors, databases, and brokers. The resource monitor is responsible for monitoring indicators of the volatile environment and MAR tasks. The broker is responsible for making scheduling decisions based onthe remaining resources and tasks. Specifically, the layer is implemented through virtual machines that can be distributed across multiple physical servers to utilize their computing resources in making scheduling decisions.\footnote{Management layer and computation layer are connected via REST API (Representational State Transfer Application Programming Interface).} In the computation layer, there are two types of servers that can provide computation: 1) Edge servers that are deployed at the end of the networks, which are close to the base stations. 2) Cloud servers that are deployed far from the base stations, which are several hops from the users. Typically, edge servers have lower latency with less computing power, while cloud servers have higher latency but are generally more computation-powerful. Tasks that are not completed in the current time interval will be rescheduled in the next time interval and migration overheads may occur during this process. The migration within the edge or cloud layer can be neglected, while the migration between the cloud layer and the edge layer will cost a lot of time and energy, and thus cannot be neglected.
 
  Due to the ever-changing number of access users and the heterogeneity of task computation, environmental parameters such as channel bandwidth and host computation resource utilization change constantly. Besides, users’ movement in geographical locations will cause regular disconnection and long handovers, which will lead to changes in devices’ latency. These explain why the resource monitor is introduced. With these constantly changing state parameters at hand, the broker is responsible for finding the optimal strategy for offloading tasks to different hosts, known as scheduling. Specific scheduling metrics and strategies will be detailed in Section \Rmnum{4}.

\begin{figure}
  \centering
  \includegraphics[width=7.5 cm]{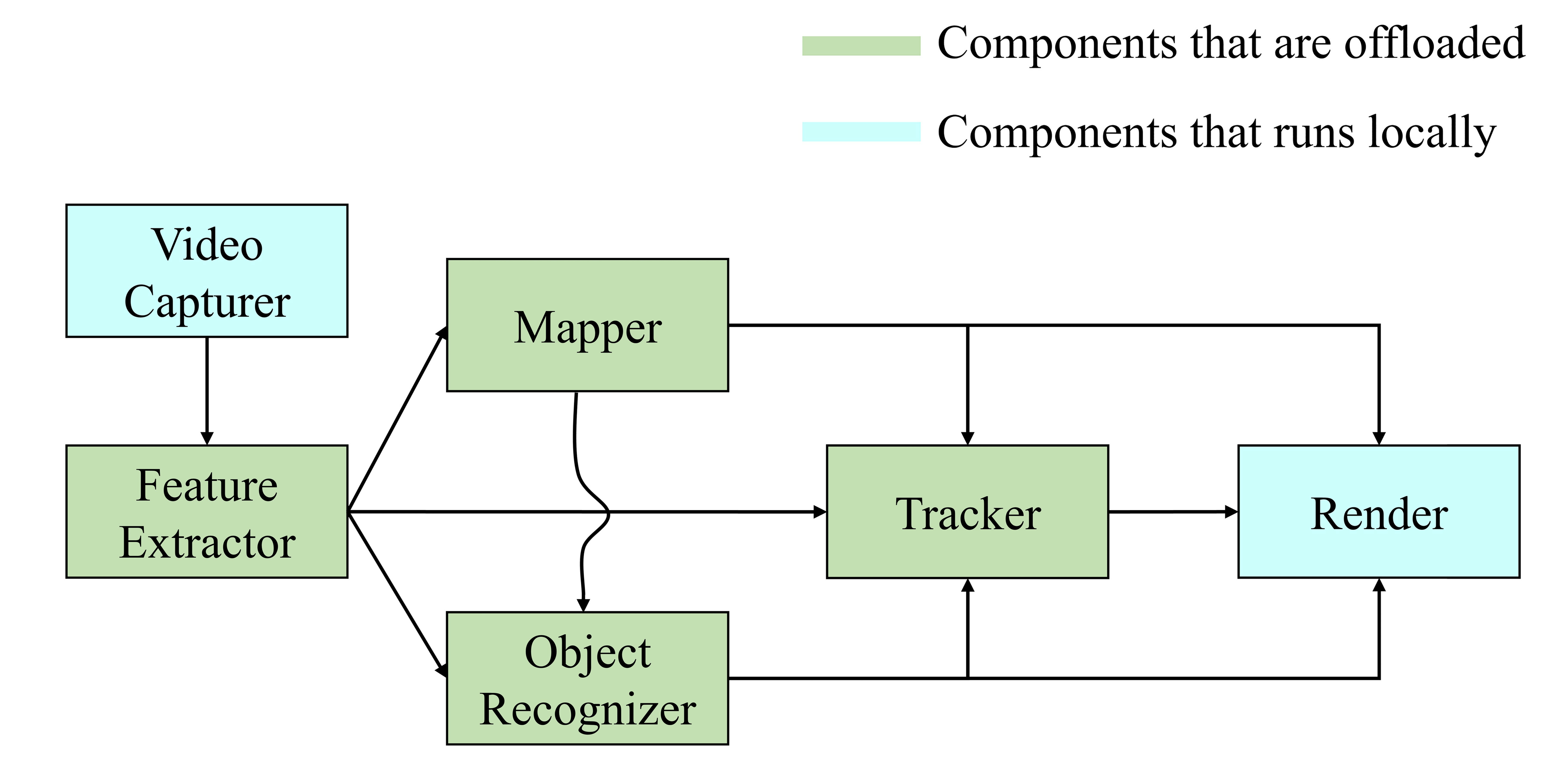}
  \caption{Main components of typical MAR application and their interdependency.} 
  \label{fig2}
\end{figure}

\section{MAR Application Model} 
In this section, we conduct modeling analysis for MAR based on the application case of web browser \cite{Lam2019}. Web browser is a typical MAR application that combines geolocation and time information with computationally intensive image processing algorithms to display location-specific websites content over original video stream. 
Instead of modeling one MAR application as a whole task \cite{Bi2018} or splitting an application into several subtasks using standalone workloads without interdependency\cite{Wu2018}, we develop a MAR workflow model that takes full account of the precedence constraints of MAR applications, as shown in Fig. \ref{fig2}. Compared with standalone workloads, the workflows in Fig. \ref{fig2} reveal the internal relationship between different MAR tasks, making the structure of the whole system clearer and easier to monitor, which facilitates the execution of the scheduling program.

The workflow starts with the mobile camera capturing the raw video frame. The raw video frame then goes through the feature extractor which extracts its feature points. These feature points will then be sent to three interdependent components, i.e., the mapper, the tracker, and the object recognizer. Using the feature points, the mapper can build a digital model of the 3-dimensional (3D) environment, which will be sent to the object recognizer together with the feature points to locate a specific object. Using the results of the object recognizer, the tracker can track the object in the next few frames. The results of these three components will then be transmitted back to the render. Finally, render will combine them together to obtain the positional and image information and overlay the virtual content on top of the original video stream for the user based on that information. For example, suppose Object Recognizer detects a restaurant in the digital model created by the mapper, the tracker can then effectively track the presence of that restaurant in the subsequent video stream. The location of the restaurant in the image is then sent to the Render component, which can use this information to overlay relevant virtual content about the restaurant from the internet on top of the video stream. This virtual content may include details about average spending at the restaurant or other people's comments, providing users with location-specific information in real-time, making the overall browsing experience more interactive and informative. The video capturer and the render are the components that directly interact with users, and thus can only be run locally, while the other components are more computationally-intensive, and are typically offloaded to the edge or cloud servers for execution.
For the mapper, we use ORB-SLAM2, which is a complete simultaneous localization and mapping (SLAM) system for monocular, stereo, and RGB-D cameras, including map reuse, loop closing, and relocalization capabilities \cite{Mur-Artal2017}. For object recognizer, we use the YOLO algorithm\footnote{https://github.com/ultralytics/yolov5}, which is responsible for identifying and locating objects in videos. Tracker refers to object tracker, we use OpenCV\footnote{https://opencv.org/} for tracking in this work, which can track the objects recognized by YOLO in several subsequent frames. 

\begin{figure}
  \centering
  \includegraphics[width=8.5cm]{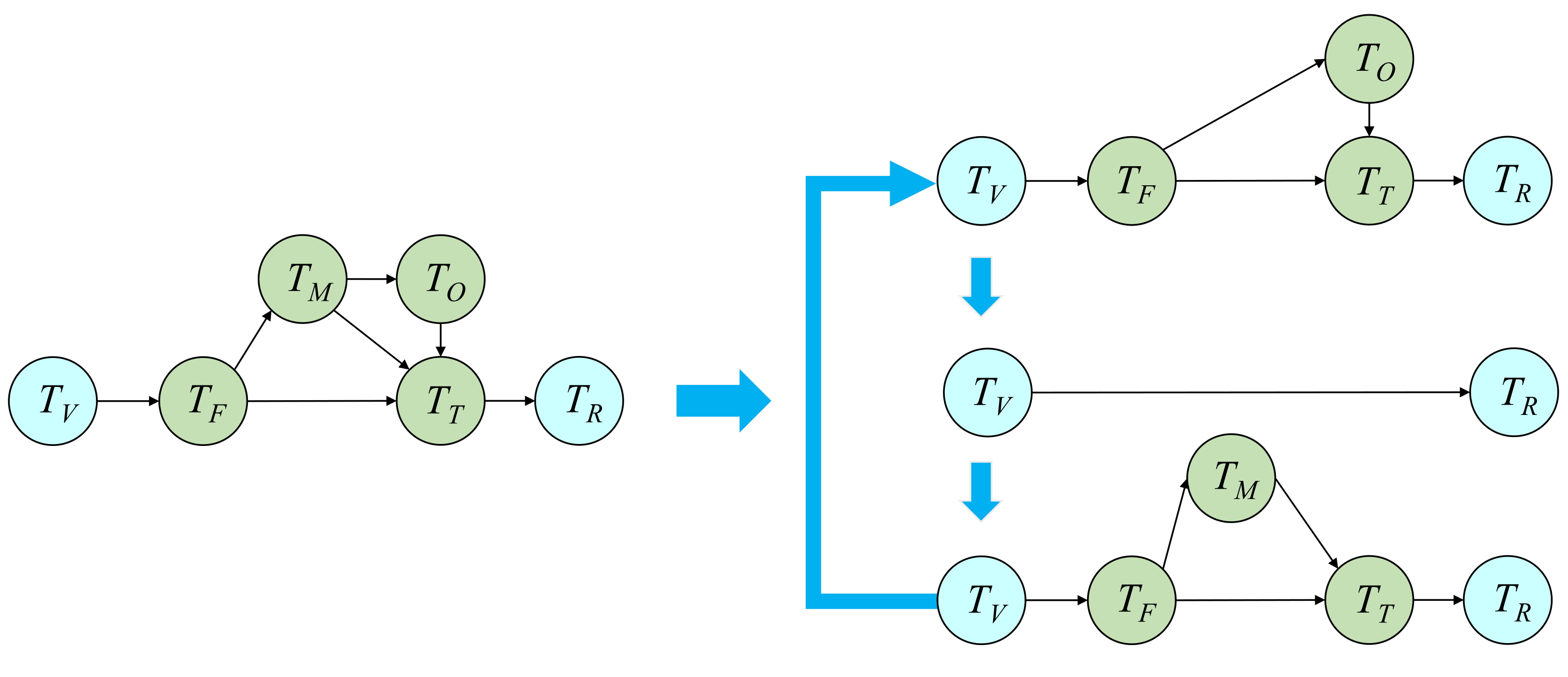}
  \caption{MAR subtask simplified into three types of DAG according to different time intervals. $T_V, T_F, T_M, T_O, T_T, T_R$ represent the video capturer, the feature extractor, the mapper, the object recognizer, the tracker, and the render, respectively.} 
  \label{fig3}
\end{figure}

In this article, we consider the frame rate of the videos captured to be $60$ frames per second, so the deadline for feature extractor $t_d = 1/60$. The render has a strict latency requirement as it needs to be executed every frame so as not to affect the user’s experience \cite{BRAUD2020}. The tracker is unnecessary to be called every frame, its deadline can be $2t_d$. Mapper runs as a background task constantly refining and expanding the 3D map and thus has no strict latency requirements, its deadline can be $3t_d$. The object recognizer also has more relaxed requirements, as delays in the order of a second before relocalizing or annotating are still sufficient to achieve acceptable user experience, its deadline can be $4t_d$\cite{BRAUD2020} \cite{Verbelen2012}. Considering their deadline requirements, this model can leverage the results from the previous frame, thus avoiding excess computation and latency without influencing user experiences. Based on the analysis of this paragraph, four directed acyclic graphs of the application are shown in Fig. \ref{fig3}.

\section{Problem Formulation} 
\begin{figure*}
  \centering
  \includegraphics[width=17.5cm]{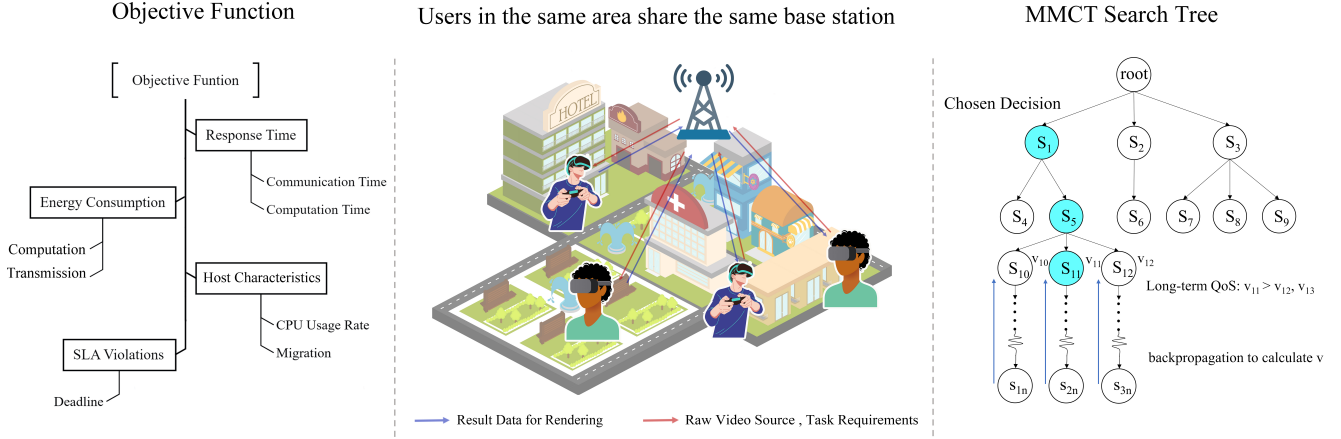}
  \caption{The objective function, a system application scenario, and the MMCT search tree} 
  \label{fig4}
\end{figure*}

After a decision is executed, its QoS can be determined quantitatively by an objective function that can be denoted as $Y$. To find the highest QoS score, we need to find the optimum schedule decision such that the objective function is minimized. This objective function consists of four first-level QoS indicators including the response time, the energy consumption, the host characteristic, and the service-level agreement (SLA) violations. With these indicators represented as $ARS$, $AEC$, $HC$, and $SLA$ respectively, the problem can be formulated as:
\begin{center}
$Y=\alpha ARS+\beta AEC+\gamma HC+\delta SLA.$
\end{center}
In this formula, the $\alpha$, $\beta$, $\gamma$, and  $\delta$ stand for their corresponding
trade-off coefficients. The coefficients can be adjusted considering the features of MAR devices and different users’ requirements.\\
 The structure of the objective function is shown in Fig. \ref{fig4}, each first-level indicator consists of several second-level indicators, which are discussed in detail as follows.\\

\subsection{Response Time}
The response time of a MAR task consists of the communication latency and computation time of each task. The computation time is only relevant to task complexity and instruction per second that the current host could provide. The communication latency includes the transmission time, the connection time, and the queuing time. The transmission time includes the transmission time of original data and results, it is relevant to the data sizes and the uplink and downlink bandwidth. The connection time refers to the response time of the edge and cloud servers, which we consider being $0.5$ $\mathrm{ms}$ and $5$ $\mathrm{ms}$, respectively. The queuing time refers to the time spent before a task is scheduled to a host.

\subsection{Energy Consumption}
In this work,  the transmission energy consumption and the computation energy consumption are considered. The transmission to the cloud server requires larger power than that to the edge server because of the geographic distance differences, thus giving rise to differences in energy consumption. The computation energy in hosts and mobile devices is proportional to the task complexity. 

\subsection{Host Characteristics }
To achieve better long-term QoS scores, it is necessary to balance the computation resource usage of each host to ensure the efficient execution of tasks. Meanwhile, when task migration transfers the tasks of a host to another host, it causes additional transmission time and also disrupts the original computing process of the host, so it has a lot of internal hidden negative effects on the efficiency. Finally, we use the variance of host central processing unit (CPU) utilization and the number of host migrations to represent host characteristics.

\subsection{SLA Violations}
SLA violation refers to the number of times tasks have not been completed within the deadlines. This metric can have a significant impact on the user experience, causing a perceptible delay, and thus need to be met in the first place.\\

When calculating the value of the objective function, all these different indicators will be normalized and given different weights according to user requirements. The value of the objective function will be the weighted summation of all indicators. 
\begin{figure*}
  \centering
  \includegraphics[width=17.5cm]{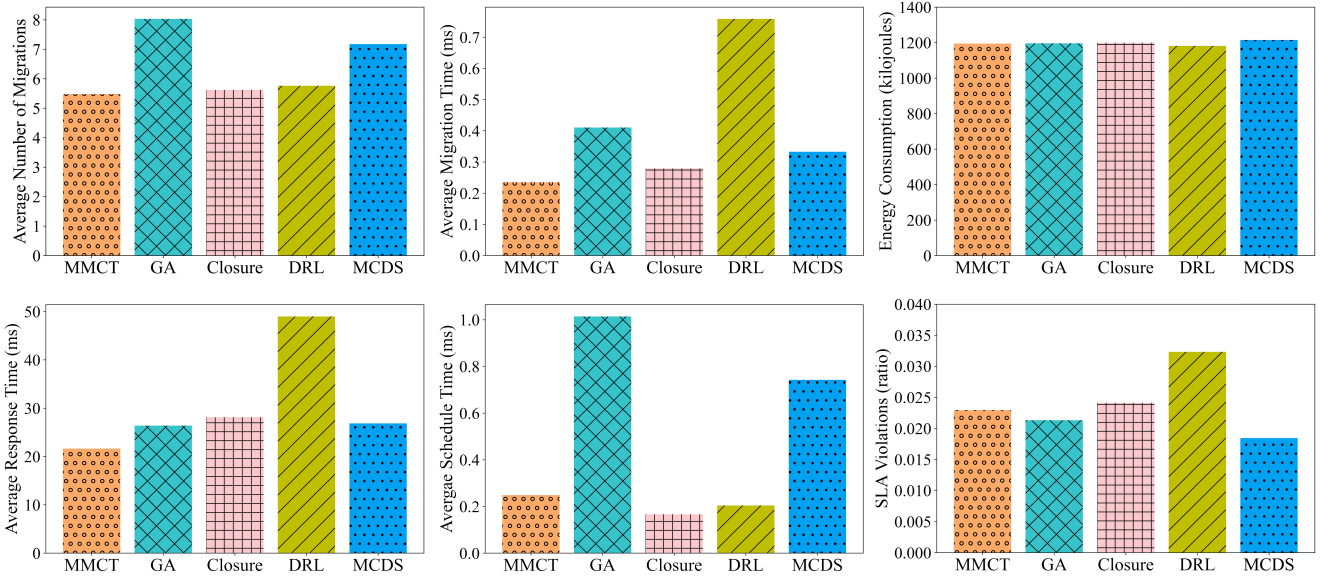}
  \caption{Numerical results of the experiments conducted in multi-user edge-cloud computing systems} 
  \label{fig5}
\end{figure*}
\section{MMCT Framework}
In this article, we propose an MMCT for look-ahead offloading decision-making. Traditional Monte Carlo tree search (MCTS) is a look-ahead scheme that runs several multi-step executions to get an unbiased estimate of the long-term effect of an immediate action \cite{Qian2022}. Generally speaking, running these look-ahead executions in a real environment is time-consuming and hence infeasible for real-time scheduling. Fortunately, a recently proposed coupled-simulation framework leverages event-driven simulators to quickly get a QoS estimate to avoid executing complex decisions on a physical platform \cite{Tuli2022}. Using these simulated results, we create an objective function as discussed above, so as to get a QoS score for each decision to estimate its effect in near real-time. What's more, to further facilitate the efficient use of computing resources, we establish a mechanism to narrow the searching scope while iteration so as to accelerate convergence. 

In each interval, several executable tasks will be sent to the management layer. Offloading a task $t_i$ to a host $h_i$ is represented as decision $(t_i, h_i)$, and all possible decisions are defined as the decision spaces. We choose a point from the decision space randomly as the root node (initial state), where the whole algorithm starts from. After the root node is selected, other points in the decision space with unscheduled tasks serve as leaf nodes.  Unlike other problems where there is seldom or no migration overhead between each decision, MAR scheduling problem has unpredictable migration overhead between decisions. For instance, a root node with a lower response time may incur a higher migration overhead in the next interval and vice versa. Therefore, relying on heuristics or prior knowledge to select the root node may result in a bias towards a specific subset of the decision space. Random selection of the root node can promote a thorough exploration of the decision space and mitigate any potential biases.

Each node can be defined with $[(t_i, h_i), q_i, v_i, n_i]$, where $q_i$, $v_i$ are initialized as $0$, $n_i$ is initialized as $1$. $q_i$, $v_i$ and $n_i$ represent the immediate QoS score, the long-term QoS score, and the number of visits to this node, respectively. 

The MMCT search consists of five steps: selection, expansion, simulation, backpropagation, and discards. Through these five steps, a search tree will be created, as shown in Fig. \ref{fig4}. The search tree will be refreshed at most $M$ times, gradually converging to the optimal leaf node, which will be taken as the offloading decision finally. Next, we detail the five steps in this algorithm.

\subsection{Selection}
In the algorithm, the leaf node with the highest upper confidence bound (UCB) will be chosen \cite{Tuli2022a}. Specifically, the selection uses the rule:
\begin{center}
\ $\mathrm{Leaf~node}= \underset{i}{\arg\ }\max\  v_i+\sqrt{\dfrac{c\times\ln n}{n_i}}$,
\end{center}
where  $n$ represents the number of visits to the root node. $c\in[0,1] $ is exploration parameter. When the $c$ is set to a larger size, the algorithm will converge to a leaf node faster. We set $c$ as 0.5 in this work.

\subsection{Expansion}
After selection, up to four child nodes can be generated from the selected leaf node using the decision in decision space to expend the searching tree. The immediate QoS score $q_i$ of these child nodes will be calculated. Then, this leaf node becomes the new root node, while the child nodes become new leaf nodes. 

\subsection{Simulation}
For the newly generated nodes, each node is simulated for $N$ steps, and each step is made by randomly selecting a single decision point of an unassigned task from the decision space, represented as $[(t_j, h_j), q_j, v_j, n_j]$, with its $q_j$ calculated and $v_j$, $n_j$ initialized as $0$ and $1$. $N$ is the number of look-ahead simulations called roll-out parameter. When it is set to a larger size, it can lead to a better decision but is more computationally expensive. 

\begin{figure*}
  \centering
  \includegraphics[width=17cm]{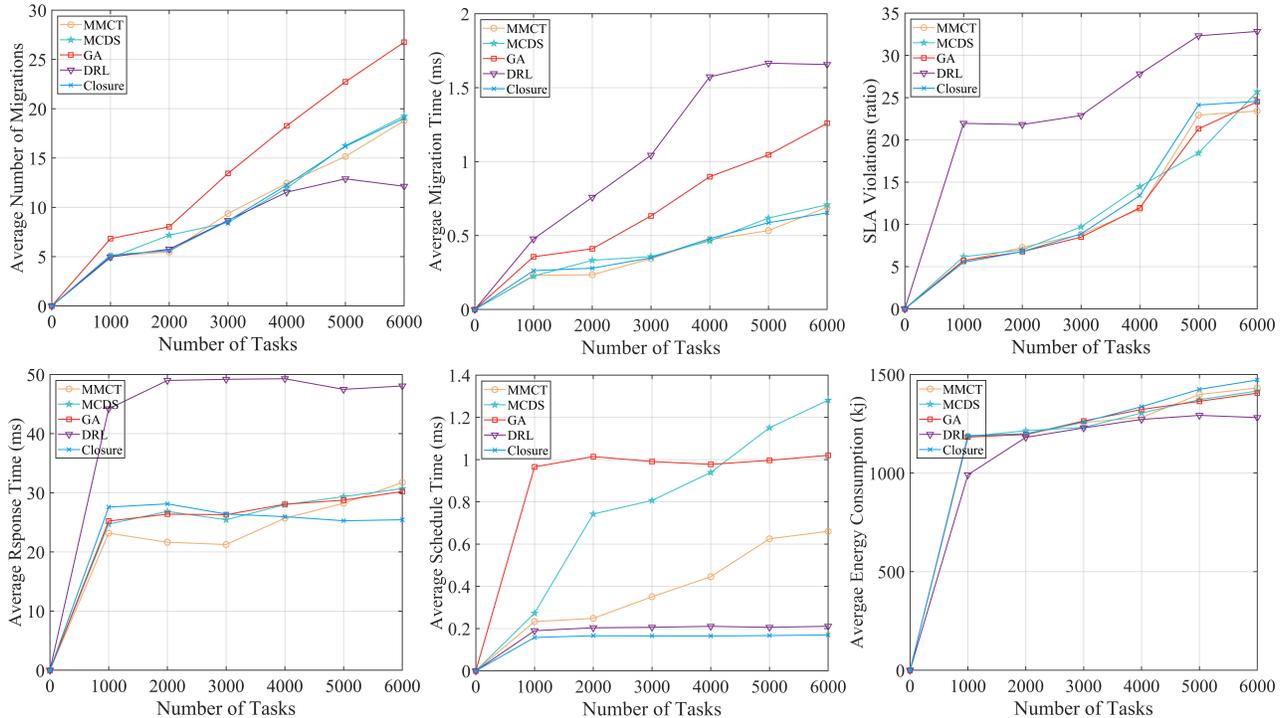}
  \caption{Numerical results of the scalability experiments conducted in edge-cloud computing system} 
  \label{fig6}
\end{figure*}

\subsection{Backpropagation}
After calculating the immediate effect $q_j$ in the future $N$ decisions of a leaf node, we can use these $q_j$ values to backpropagate from the last node step by step to get the long-term QoS score of the current leaf node. This is a weighted summation process. The farther away from the leaf node indicates the lower the weight of the simulated node fraction.

\subsection{Discard}
In the traditional MCTS algorithm, the above four processes will be repeated $M$ times, and the leaf node with the highest number of visits will be chosen as final decision in the end. However, in practice, when the number of leaf nodes is large and the objective function values of each node differs greatly, the nodes that show poor potential can be discarded, so as to narrow the search area. Specifically, we consider that when the number of visits of a leaf node $n$ is greater than $\left[M/2\right]$, there won't be a leaf node with higher $n$ in the search tree, so the rest of the tree can be discarded to save computing resources. 

\section{Performance Evaluation}
In the experiments, we simulate a large-scale multi-user edge-cloud computing system with $30$ edge hosts and $20$ cloud hosts in total using Microsoft Azure virtual machines (VMs)\footnote{Azure general purpose B-type VMs:https://docs.microsoft.com/en-us/azure/virtual-machines/sizes-b-series-burstable}. Specifically, we use the Azure B2s and B8ms machines as edge servers and cloud servers, respectively. The Azure B2s machines consist of two cores (4029 MIPS), and B8ms machines have eight cores (1601 MIPS). The experiments are conducted in a simulated MAR scenario with MAR tasks generated every time interval. In the experiment, the number of look-ahead simulations $N$ is set to be $7$ and the maximum number of iterations $M$ is set to be $10$. They are determined based on a grid search that evaluated the convergence and stability of the algorithm under different values. 
For comparison, we also simulate the GA, Closure \cite{Wang2020}, DRL \cite{Chen2021}, MCDS \cite{Tuli2022a} algorithms-based offloading. We compare these schemes in the average number of migrations, the average migration time, the energy consumption, the average response time, the average schedule time, and the SLA violations, the results with a fixed number of $2000$ tasks are shown in Fig. \ref{fig5}.

Overall, the results indicate that MMCT model can effectively promote the QoS compared to the other four benchmark schemes. Specifically, MMCT has the relatively lowest response time and migration time, other indicators like SLA violations and scheduling time are within the acceptable range.

It can be seen that although both MCDS and MMCT are based on the Monte Carlo tree search, the performance of MMCT is much better than MCDS, especially in scheduling time. This is mainly because our proposed model takes into account the volatile and dynamic nature of the multi-user MAR environment, allowing for more adaptive and intelligent decision-making. Energy consumption in Fig. \ref{fig5} represents the energy consumed by the whole edge-cloud system, where the energy consumed by the mobile device is only relevant to the data size of the task and the location of the user.

Besides, it is noticed that MMCT can reduce migration overheads significantly. As shown in Fig. \ref{fig5}, MMCT outperforms the other four schemes in terms of the average migration time and the average number of migrations, especially over single-step scheduling schemes like GA and DRL-based algorithms. Two factors contribute to this positive outcome.  Firstly, the look-ahead ability of MMCT allows it to estimate the long-term effects of decisions and prevent potential significant migration overhead. Secondly, the workflow model takes into account the interdependency of MAR applications, enabling more comprehensive and accurate predictions of task execution time and resource requirements. On the contrary, single-step offloading like GA and DRL-based schemes are rather myopic, creating a lot of non-local jumps in the search space while offloading. 

In addition, we also test and compare the scalability of the above-mentioned schemes, we change the number of average workflows (access users) in each time interval and observe the changes in each metric, the results are shown in Fig. \ref{fig6}. MMCT has shown better scalability compared with the other four schemes. Results indicate that all metrics of MMCT have been maintained at a relatively good level when the number of users grows. The intuitive reason is that MMCT considers several tasks as a whole while offloading and searches and compares possible decisions before offloading. In this way, MMCT can make better use of limited computation resources and thus is more stable when the access of users grows.

\section{Open Research Directions and Challenges}

To better provision multi-user MAR services in edge-cloud computing networks, there are several issues to be further addressed.
\subsection	{Motion-Aware Task Offloading}
A motion-aware scheduler can select keyframes from the video source \cite{Ren2022}. By only offloading these keyframes to the back-end server for feature extraction and object recognition, the computational efficiency can be further improved. However, due to the loss of feature points caused by the tracking algorithm and the inherent difficulty of recognizing the user’s movement behavior based on the isolated video frame, this technology is still immature.

\subsection	{Preloading for Virtual Content }	
By modeling the user devices’ motion trajectories as Markov decision process (MDP) and adaptively learning the optimal preloading policy, AR intelligent preloading algorithm can proactively transmit holographic contents to the devices \cite{Han2022}. However, due to the diversity of user devices’ motion preferences, the edge server needs to provide individual preloading solutions for each user device, which would lead to high computation complexity when the number of user devices increases.

\subsection	{MAR Content Sharing}
One of the key observations in MAR is that nearby AR users may share some common interests, and may even have overlapped views to augment. By leveraging this feature, MAR devices can share the content among themselves and reuse the content, which helps relieve the edge workload and address the scalability issue. The tricky part of this system is to determine with whom to share the recognition results that are stored in the local cache and how to address the privacy issue when it comes to certain scenes\cite{Zhang2022}.

\section{Conclusion}
In this article, we first established three types of workflow models for MAR applications according to different time interval periods. Then we proposed an MMCT search method, which can efficiently schedule workflows for MAR applications. 
Compared with the existing MAR scheduling schemes,  the MAR workflow-oriented look-ahead scheduling method can find the optimal scheduling scheme for the long-term QoS, rather than being myopic by scheduling a single task at a time. Moreover, the migration overhead was considered in workflow scheduling. The methods are more relaxed about environmental assumptions and thus can adapt to more volatile environments.

\bibliographystyle{IEEEtran} 
\bibliography{IEEEabrv,bib}

\textbf
{Ruxiao Chen} is currently a junior student at Shandong University. He is also an active participant in the seminar of the Shandong Key Laboratory of Wireless Communication Technologies. Despite being early in his academic journey, he has already made several contributions to the field. he has engaged in research projects and discussions with peers and professors, always seeking to expand his knowledge and hone his skills. His research interest lies in exploring the intersection of AI and edge computing, as well as how machine learning can be leveraged to solve complex problems in real-world settings.

\textbf
{Shuaishuai Guo}(Senior Member, IEEE) received the B.E and Ph.D. degrees in communication and information systems from the School of Information Science and Engineering, Shandong University, Jinan, China, in 2011 and 2017, respectively. He visited the University of Tennessee at Chattanooga (UTC), USA, from 2016 to 2017. He worked as a postdoctoral research fellow at King Abdullah University of Science and Technology (KAUST), Saudi Arabia from 2017 to 2019. Now, he is working as a full professor of Shandong University. His research interests include 6G communications and machine learning.

\end{document}